\documentclass[prc,aps,nofootinbib,twocolumn]{revtex4-1}
\usepackage[utf8]{inputenc}
\usepackage{mathrsfs}
\usepackage{amssymb}
\usepackage{amsmath}
\usepackage{graphicx}

\usepackage[caption = false]{subfig}
\usepackage[normalem]{ulem}
\usepackage{xcolor}
\usepackage{hyperref}
\usepackage{slashed}
\usepackage{tikz}
\usepackage{cleveref}
\usepackage{xspace}

\newcommand{\dd}{\mathrm{d}}
\newcommand{\ee}{\mathrm{e}}

\DeclareMathOperator{\BessK}{K}

\newcommand{\UNIT}[1]{\ensuremath{\,{\rm #1}}\xspace}

\newcommand{\MeV}{\UNIT{MeV}}
\newcommand{\GeV}{\UNIT{GeV}}

\newcommand{\fm}{\UNIT{fm}}
\newcommand{\mb}{\UNIT{mb}}
\newcommand{\proz}{\UNIT{\%}}

\newcommand{\REM}[1]{}

\definecolor{magenta}{cmyk}{0,1,0,0}

\makeindex

\begin{document}

\title{The Shear Viscosity to Entropy Density Ratio of Hagedorn States}

\author{Jan Rais}
\email{rais@th.physik.uni-frankfurt.de}
\affiliation{Institut f\"ur Theoretische Physik,
  Johann Wolfgang Goethe-Universit\"at,
  Max-von-Laue-Str.\ 1, D-60438 Frankfurt am Main, Germany}

\author{Kai Gallmeister}
\affiliation{Institut f\"ur Theoretische Physik,
  Johann Wolfgang Goethe-Universit\"at,
  Max-von-Laue-Str.\ 1, D-60438 Frankfurt am Main, Germany}

\author{Carsten Greiner}
\affiliation{Institut f\"ur Theoretische Physik,
  Johann Wolfgang Goethe-Universit\"at,
  Max-von-Laue-Str.\ 1, D-60438 Frankfurt am Main, Germany}

\date{\today}


\begin{abstract}
  The fireball concept of Rolf Hagedorn, developed in the 1960's, is an alternative description of hadronic matter.
  Using a recently derived mass spectrum, we use the transport model GiBUU to calculate the shear viscosity of a gas of such Hagedorn states, applying the Green-Kubo method to Monte-Carlo calculations.
  Since the entropy density is rising ad infinitum near $T_H$, this leads to a very low shear viscosity to entropy density ratio near $T_H$.
  Further, by comparing our results with analytic expressions, we find a nice extrapolation behavior, indicating that a gas of Hagedorn states comes close or even below the boundary $1/4\pi$ from AdS-CFT.

\end{abstract}

\maketitle


\section{Introduction}
\label{sec:Intro}

The properties of hot and dense matter, created experimentally in heavy-ion collision performed at accelerators like RHIC or CERN, are usually extracted by applying relativistic hydrodynamics or kinetic transport theory.
Doing hydrodynamics, transport coefficients like heat or electric conductivity, or shear- or bulk viscosity, are extrinsic inputs which should be calculated from an underlying field theory, as it is Quantum Chromodynamics (QCD) for the for the quark gluon plasma (QGP).
The shear viscosity, as one of those transport coefficients, can be calculated employing two-particle scattering processes. Dealing with QGP, there is almost a perfect liquid characterized by a very small value for the shear viscosity to entropy density ratio, $\eta/s$.
Nevertheless, this ratio never undergoes the value $1/4\pi$, which is derived within the anti-de Sitter/conformal field theory (AdS/CFT) \cite{Kovtun:2004de}. This boundary holds for all substances in nature.

In \cite{Xu:2007ns} it was shown, within the BAMPS parton cascade, which includes inelastic gluonic $gg \leftrightarrow ggg$ reactions, that $\eta/s \sim 0.13$ in a pure gluon gas. This is as expected, because $\eta/s$ increases with decreasing $T$, which goes hand in hand with a decrease of the relevant hadronic cross section in the hadronic phase \cite{Gavin:1985ph,Venugopalan:1994ux}. On the other hand, asymptotic freedom dictates that $\eta/s$ increases with $T$ in the deconfined phase. Here the coupling between quarks and gluons decreases logarithmically \cite{Arnold:2003zc}.

There have been several efforts to study this transport coefficient $\eta/s$ in microscopic models using the Green-Kubo formalism, as e.g.~in UrQMD \cite{Muronga:2003tb,Demir:2008tr}, in SMASH \cite{Rose:2017bjz} and in pHSD \cite{Ozvenchuk:2012kh}.
On the partonic side, either pHSD \cite{Ozvenchuk:2012kh}, PCM \cite{Fuini:2010xz}, and BAMPS have been used \cite{Wesp:2011yy},  while within the latter model also a critical test of the Green-Kubo method itself has been performed \cite{Reining:2011xn}.
Very recently, there was an attempt using a $S$-matrix based Hadron Resonance Model via the Chapman-Enskog method \cite{Dash:2019zwq}.

Before QCD made the calculation of phase transition and QGP possible, an alternative theory describing hadrons was devoloped by Rolf Hagedorn in the 1960's \cite{Hagedorn:1965st}. He states a visual concept that reads
  "fireballs, consist of fireballs, which consist of fireballs \dots".
This yields a density of (hadronic) states as function of the mass as
\begin{align}\label{eq:fitfunc}
\rho(m) = \text{const.}\, m^{-a} \, \exp \left[m/T_H\right]\ ,
\end{align}
with $T_H$ being the so-called ``Hagedorn temperature''.
Later, Frautschi invented a reformulation \cite{Frautschi:1971ij}, yielding a bootstrap equation,
\begin{align}
\label{eq:bootstrap}
\rho(m) =& \rho_0(m)\ +\ \sum_N\frac{1}{N!}\left[\frac{V_0}{(2\pi)^3}\right]^{N-1}\\
&\times \int \prod_{i=1}^{N}\left[\dd m_i \rho(m_i) \dd^3p_i\right] \delta^{(4)}\left(\sum_ip_i-p\right)\ .\nonumber
\end{align}
Here $V_0$ is the volume of the Hagedorn states.
In general, this equation can not be solved analytically. For the easiest inhomogeneity, $\rho_0(m)=\delta(m-m_0)$ with $m_0$ the mass of some initial state,
Nahm \cite{Nahm:1972zc} found a solution with $a \approx 3$.
With $V_0 \simeq (4\pi/3)m_\pi^{-3}$, $m_0 \simeq m_\pi$, one achieves a slope $T_H \simeq 150\MeV$.

For more realistic inhomogenities of \cref{eq:bootstrap}, the solution has to be found numerically.
Recently, our group developed a prescription with $N=2$, where the quantum numbers $B$ (baryon number), $S$ (strangeness) and $Q$ (electric charge) are conserved explicitly \cite{Beitel:2014kza,Beitel:2016ghw,Gallmeister:2017ths}.
Summing over all quantum numbers, one gets for the prescription given in \cite{Gallmeister:2017ths} a Hagedorn spectrum, \cref{eq:fitfunc}, which is characterized by $T_H=165\MeV$ and $a=2.98$ (for a Hagedorn state radius $R=1.0\fm$)\footnote{Non-vanishing chemical potentials disallow the independent summation over quantum numbers in the Hagedorn spectrum and the thermal distribution  \cite{Gallmeister:2017ths}}.
As will be discussed below, these numbers can only be used for analytic estimates with restrictions. For real calculations the detailed, tabulated spectra are used. Nevertheless, the fitted distribution may be used to estimate some quantities in the vicinity of $T_H$.

The aim of this paper is to study how the shear viscosity over entropy density of a Hagedorn gas behaves as a function of the temperature of the gas (cf.~also \cite{NoronhaHostler:2008ju}). Thus analytical estimates are compared to Monte Carlo results obtained from box calculations based on the Green-Kubo formalism. In order to check the validity of the results, also results for a pion gas are analyzed, where three different charge states may interact elastically according an isotropic constant cross section $\sigma=30\mb$. Thus the interaction is direct comparable to that of the Hagedorn gas.

The paper is organized as follows.
In section II, analytic expressions for the thermodynamical quantities of the considered Hagedorns state gas are given.
Also, an analytic expression for the shear viscosity of a gas of particles, which interpolates the non-relativistic regime to the relativistic regime necessary for pions is stated.
Section III describes the numerical Green-Kubo method used in this analysis and shows intermediate results.
The final results for $\eta$ and $\eta/s$ are presented in section IV and discussed in section V.
%

\section{Analytic Expressions}
\label{sec:AnaExpr}

\subsection{Thermodynamical quantities}
\label{sec:thermoQuantities}

Knowing the general resonance gas partition function in Boltzmann approximation,
\begin{align}\label{eq:partitionfunction}
\ln \mathcal{Z}(T,V) = \frac{VT}{2\pi^2} \int \dd m \, \rho(m)\,m^2 \BessK_2\left(\frac{m}{T}\right)\ ,
\end{align}
with $\BessK_\nu$ being a modified Bessel function, one may derive all necessary thermodynamical quantities (cf.~e.g.~\cite{NoronhaHostler:2012ug}), as e.g.~particle density $n$, energy density $e$, and entropy density $s$, as
\begin{align}
  \label{eq:nDens}
  n(T)&=\frac{T}{2\pi^2}\int \dd m \, \rho(m)\,m^2\BessK_2\left(\frac{m}{T}\right)\ ,\\
  \label{eq:eDens}
  e(T)&=\frac{T}{2\pi^2}\int \dd m \, \rho(m)\,m^3\left[3\frac{T}{m}\BessK_2\left(\frac{m}{T}\right)+\BessK_1\left(\frac{m}{T}\right)\right]\ ,\\
  \label{eq:sDens}
  s(T)&=\frac{1}{2\pi^2}\int \dd m \,\rho(m)\,m^3 \BessK_3\left(\frac{m}{T}\right)\ ,
\end{align}
where all chemical potentials have been neglected, $\mu=0$.
Since for Boltzmann statistics the pressure is given by $p=T\,n$, it is easily observed using the recurrence relations of the Bessel functions, that the well known Gibbs-Duhem relation,
\begin{align}\label{eq:sHag}
s &= \frac{e + p}{T} \ ,
\end{align}
is fulfilled.

\begin{figure}[htb]
  \centering
  \includegraphics[width=\columnwidth]{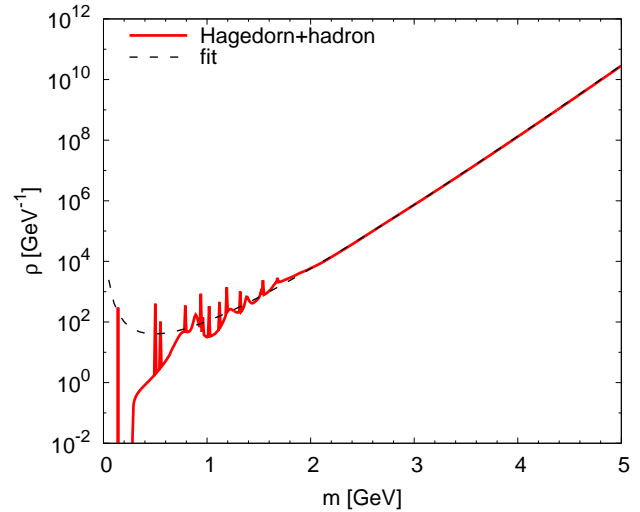}
  \caption{The Hagedorn spectrum and a fit according \cref{eq:fitfunc} ($T_H=0.165\GeV$, $a=2.98$) as function of the mass $m$.}
  \label{fig:HagSpectrum}
\end{figure}

It is important to note, that a simple minded insertion of \cref{eq:fitfunc} into the thermodynamical integrals \cref{eq:nDens,eq:eDens,eq:sDens} leads to faulty results: The Hagedorn spectrum fitted to a function according \cref{eq:fitfunc} only describes the high mass ($m\gtrsim 2\GeV$) contribution, but totally fails below. Here the full Hagedorn spectrum is a sum of the known hadron states and the pure Hagedorn states, thus showing all the mass structures of the known hadrons, \cref{fig:HagSpectrum}.
Therefore we will use the tabulated spectrum of hadrons and Hagedorn states instead of an analytic approximation in all what follows, except the extrapolations described below. The tabulation stops at Hagedorn state masses $m=10\GeV$.

\begin{figure}[htb]
  \centering
  \includegraphics[width=\columnwidth]{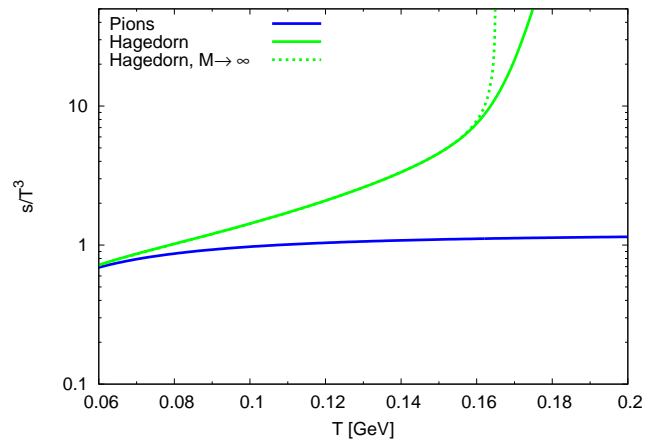}
  \caption{The entropy density $s$, \cref{eq:sDens}, normalized to $T^3$ as function of the temperature $T$. Also the estimate for $m\to\infty$ (see text for details) is shown.}
  \label{fig:s_T3}
\end{figure}

The resulting entropy as a function of temperature is shown in \cref{fig:s_T3}.
The entropy density for pions is simply calculated by replacing $\rho(m)$ by a properly scaled delta peak according the degeneracy at the pion mass.
While the entropy density of the pion gas increases very slowly, the entropy of the Hagedorn gas increases exponentially and gets very steep for $T\gtrsim T_H$. Since the Hagedorn state tabulations only extends up to masses $m=10\GeV$, the (expected) divergence is weakened, showing only a constant increase on a logarithmic scale. This holds true for all thermodynamical quantities mentioned above, as e.g.~energy and particle density.
For some quantities, it is now possible to add the missing contribution by using the analytic fit function \cref{eq:fitfunc} getting the real divergence. Inserting approximations for the Bessel function $K_\nu$ for large arguments, integrals like e.g.~\cref{eq:nDens,eq:eDens,eq:sDens} may be expressed in terms of the complementary incomplete gamma function. The corresponding result for the entropy density is also shown in \cref{fig:s_T3}.

It is obvious that one has to abstain from this procedure, when directly comparing to the Monte Carlo simulations.

We have to mention, that we consider the gas particles to be pointlike, such that there is no volume correction. Since the Hagedorn spectrum generates more and more particles, this also influences the space in a given box volume. Therefore it would be instructive to introduce volume corrections, as e.g.~in \cite{NoronhaHostler:2012ug,Rischke:1991ke,Gorenstein:2007mw} in future studies.

\subsection{Shear viscosity}
\label{sec:shearviscosity}

To investigate the shear viscosity of pion or Hagedorn states gas, it is important to ensure that the underlying formulae are valid for the desired range of the variable $z=m/T$, while $m$ is the mass of the particle and $T$ the temperature of the system. For relevant temperatures $T=100-200\MeV$ and masses $m>138\MeV$, the covered range is $z=10^{-3}-1.5$. Thus one needs a non-relativistic prescription, which reaches till $m\sim T$. For this the expression valid for all masses and all temperatures is selected as \cite{DeGroot:1980dk}
\begin{align}\label{eq:eta_general}
  \eta=\frac{15}{16}\,\frac{T}{\sigma}
  \frac{z^4\,\BessK_3^2(z)}{(az^2+b)\BessK_2(2z)+(cz^3+dz)\BessK_3(2z)}\ .
\end{align}
Choosing the values of the constants as
\begin{align}
  a=15\,,\ b=2\,,\ c=3\,,\ d=49\,,
\end{align}
yields the well known first order approximations \cite{DeGroot:1980dk}\footnote{Please note the typo concerning $a$ in the original references \cite{ANDERSON1977408,DeGroot:1980dk}. The prefactor is chosen here such that $\sigma=\sigma_{\rm tot}$.}. By slightly adjusting these constants to
\begin{align}\label{eq:eta_FaksMod}
  a=14.55\,,\ b=1.13\,,\ c=2.95\,,\ d=46.85\,,
\end{align}
\cref{eq:eta_general} gives a nice interpolation of numerical results \cite{KOX1976155} and yields the also well known higher order limiting formulae \cite{Huovinen:2008te,Wiranata:2012br}
\begin{align}
  \eta\ \underset{m\ll T}{\simeq}\ & 1.2654\,\frac{T}{\sigma}\ ,\\
  \eta\ \underset{m\gg T}{\simeq}\ & 0.3175\sqrt{\pi}\,\frac{T}{\sigma}\,\sqrt{\frac{m}{T}}\left(1+1.6349\frac{T}{m}\right)\ .
\end{align}
Thus, \cref{eq:eta_general} with the modified factors \cref{eq:eta_FaksMod} will be used further-on in this work.

Another expression covering all values of $z$ may be found in \cite{Gorenstein:2007mw}. This expression differs from the given one by more than 20\proz for $z>0.1$ and is therefore not covered here.

The overall shear viscosity of a mixture of particles is given by the weighted sum of the viscosities of each particle species \cite{Reif:1987},
which in the given case of the Hagedorn gas converts into a integration over all masses,
\begin{align}\label{eq:etachapens}
  \eta(T)&=\frac{T}{2\pi^2\,n(T)}\int \dd m \, \rho(m)\,m^2\BessK_2\left(\frac{m}{T}\right)\,\eta(m)
\end{align}
with $\eta(m)$ given by \cref{eq:eta_general,eq:eta_FaksMod}.

\section{Numerical Contemplation}
\label{sec:NumCalc}

\subsection{Implementation into GiBUU}
\label{sec:GiBUU}
The Gießen Boltzmann-Uehling-Uhlenbeck (GiBUU) project \cite{Buss:2011mx} simulates nuclear reactions as $e+A$, $\gamma + A$, $\nu+A$, $\text{hadron} + A$ (i.e.~$p+A$, $\pi+A$) or $A+A$ at energies of $10\MeV$ to $100\GeV$. Here the BUU equation
\begin{equation}
\left[\partial_t \left(\nabla_p \mathcal{H}_i \right) \nabla_r - \left(\nabla_r \mathcal{H}_i\right)\nabla_p\right] f_i(r,p,t) = C\left[f_i, f_j,...\right]
\end{equation}
is solved, where $i = N, \Delta, \pi, \rho,...$. The collision term $C$ conventionally involves the decay and scattering of 1-, 2- and 3- body processes, $C=C_{1\rightarrow x}+C_{2\rightarrow x}+C_{3\rightarrow x}$, which splits into a resonance model for low energies and the string model for high energies.
In the actual implementation \cite{Gallmeister:2017ths}, all interactions (even elastic scattering) are replaced by Hagedorn state creation and decay processes, i.e.~by $2\to1$ and $1\to2$ processes alone.
The Hagedorn spectrum tabulation limits the available energy range to be below $10\GeV$.

In the simulations, the particles are thermally initialized in a box (non-reflecting boundaries) with fixed volume of $(10\fm)^3$.
The interaction is according a constant cross section of $\sigma = 30\mb$ for the pion gas and $\sigma=\pi R^2=31.4\mb$ for the Hagedorn gas.
For the pion gas a time step size of $\Delta t=0.1\fm$ and $N_t=30000$ timesteps was chosen, while the Hagedorn gas where calculated at a lower time step size of $\Delta t=0.01\fm$ and $N_t=25000$ timesteps, which is justified because of the less steady correlation function at higher time, however bypassing too long calculation times.
These values are extracted from the error estimation via the colored noise studies described below.

It is checked, that detailed balance is fully respected and the mass and quantum number distributions are constant over the full simulation time.

\subsection{Green-Kubo formalism}
\label{sec:Green-Kubo}
The Green-Kubo method is the common method to compute transport coefficients like shear viscosity, electric or heat conductivity etc.~assuming, that the probability distribution of the time-averaged dissipative flux is Gaussian \cite{Searles:2000}. It can be derived from the dissipation-fluctuation theorem \cite{Kubo:1966,Nyquist:1928zz} and reads (see e.g.~\cite{Wesp:2011yy})
\begin{equation}
  \eta
  = \frac{1}{T} \int_V\dd^3r\int_{0}^{\infty} \dd t \left\langle \pi^{xy}(\vec r,t)\pi^{xy}(0,0)\right\rangle\ .
\end{equation}
Here $\pi^{xy}$ indicates a fixed spatial component of the volume averaged shear tensor\footnote{We use internally the three combinations $\pi^{xy}$,$\pi^{xz}$,$\pi^{yz}$ as an additional possibility to estimate the statistical error.} and $\left\langle...\right\rangle$ denotes the ensemble average of the argument.
The shear stress component, defined as
\begin{align}\label{eq:pimunucont}
  \pi^{xy}(\vec r,t)
  =\int\frac{g\dd^3 p}{(2\pi)^3\,E}\,p^xp^y\,f(\vec r,t;\vec p\,)\ ,
\end{align}
is in the simulation replaced by a discretized version,
\begin{equation}\label{eq:pimunu}
  \overline{\pi}^{xy}
  = \frac{1}{V} \sum_{i=1}^{N_{\text{part}}}\frac{p_i^x p_i^y}{p_i^0}\ ,
\end{equation}
summing up all particles in the box with volume $V$ at a given time $t$.
The correlator is obtained by the time and ensemble average in the limit $t_{\rm corr} \rightarrow \infty$,
\begin{align}\label{eq:C0_Fourier}
  C^{xy}(t)
  &=\left\langle \overline{\pi}^{xy}(t)\,\overline{\pi}^{xy}(0)\right\rangle\nonumber\\
  &=\left\langle \frac{1}{t_{\rm corr}} \int_{0}^{t_{\rm corr}} \overline{\pi}^{xy}(t+t')\,\overline{\pi}^{xy}(t')\,\dd t'\right\rangle\nonumber\\
  &=\left\langle \frac{1}{N_{\rm corr}} \sum_{j=0}^{N_{\rm corr}-1} \overline{\pi}^{xy}(i\Delta t + j\Delta t)\,\overline{\pi}^{xy}(j\Delta t)\right\rangle\nonumber\\
  &=\mathcal{F}_\omega \left[\vert \overline{\pi}_\omega^{xy}\vert^2\right](t)
\end{align}
where $N_{\rm corr} = t_{\rm corr}/\Delta t$ and $i = t/\Delta t$ and $ \mathcal{F}_\omega$ denotes the Fourier-transformed of its argument, applying the Wiener-Khinchin theorem. Here, $\overline\pi_\omega$ stands for the Fourier-transformed of $\overline\pi$.
If the system fluctuates around the equilibrium state, one finds \cite{Muronga:2003tb}
\begin{equation}
C^{xy}(t) = C^{xy}(0)\, \ee^{-\frac{t}{\tau}}.
\end{equation}
Therefore one obtains
\begin{align}\label{eq:eta}
\eta = \frac{V}{T} \int_{0}^{\infty} \dd t \, C^{xy}(t) = \frac{C^{xy}(0) V \tau}{T}\ .
\end{align}

\begin{figure}[htb]
  \centering
  \includegraphics[width=\columnwidth]{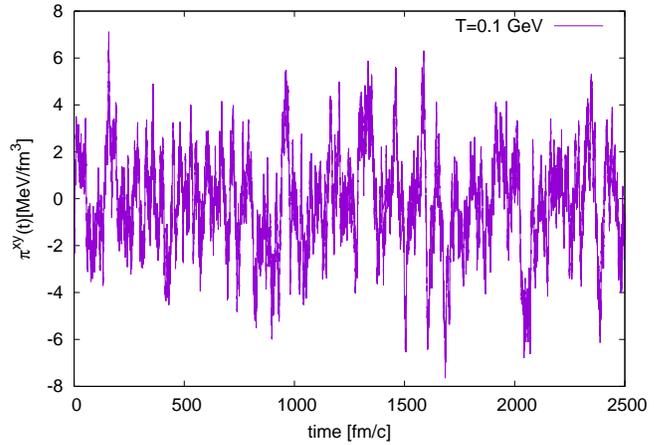}
  \caption{Example of $\pi^{xy}(t)$ as function of time $t$.}
  \label{fig:Tmunu}
\end{figure}

\begin{figure}[htb]
  \centering
  \includegraphics[width=\columnwidth]{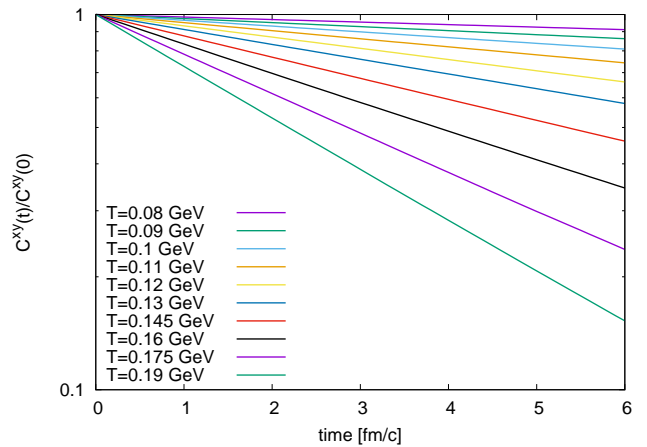}
  \caption{The correlation function $C^{xy}$ for different temperatures}
  \label{fig:slopes}
\end{figure}

This procedure is illustrated in \cref{fig:Tmunu}, showing an example of the oscillating $\pi^{xy}(t)$, and in \cref{fig:slopes}, where the exponential decaying slopes are cleary visible.

\begin{figure}[htb]
  \centering
  \includegraphics[width=\columnwidth]{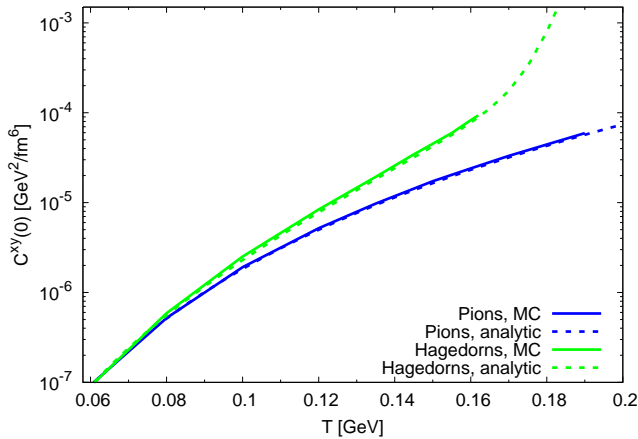}
  \caption{A comparison of analytical, \cref{eq:C0_ana}, and numerical results, \cref{eq:C0_Fourier}, of $C^{xy}(0)$.}
  \label{fig:c0}
\end{figure}

The value $C^{xy}(0)$ is of special interest because the analytic expression can be calculated easily noticing, that $C^{xy}(0) = \text{Var} \left[\overline{\pi}^{xy}\right]$. Thus, using the continuous formulation \cref{eq:pimunucont}, one obtains for one single particle species with mass $m$ and degeneracy $g$ \cite{Wesp:2011yy,Rose:2017bjz}
\begin{align}
C^{xy}_m(0)
&=\frac{g}{30\pi^2 V}\int_0^\infty \dd p \frac{p^6}{E^2} \exp\left(-\frac{E}{T}\right)
\end{align}
with $E = \sqrt{m^2+p^2}$. This integral has to be performed numerically. Finally, to get a result for the Hagedorn gas, one has to sum over all masses,
\begin{align}\label{eq:C0_ana}
C^{xy}(0) &= \frac{1}{30\pi^2 V}\int\dd m\,\rho(m)\int_0^\infty \dd p \frac{p^6}{E^2} \exp\left(-\frac{E}{T}\right)\ .
\end{align}
Irrespective of the numerical integrations, we will call these results still 'analytical' to contrast them from the results obtained via the Monte Carlo calculations.
One observes a very nice agreement of analytical, \cref{eq:C0_ana}, and numerical results, \cref{eq:C0_Fourier}, as shown in \cref{fig:c0}.

\begin{figure}[htb]
  \centering
  \includegraphics[width=\columnwidth]{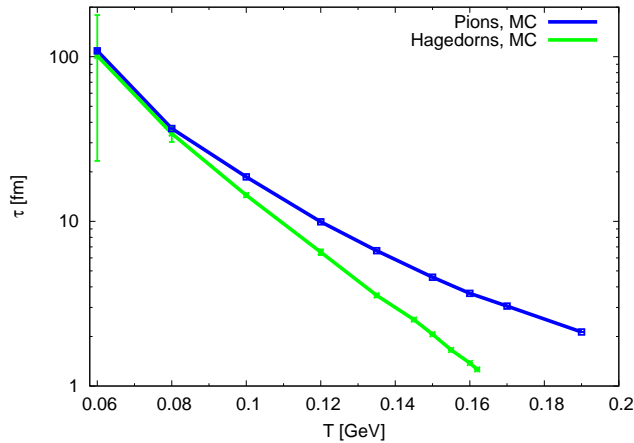}
  \caption{Results for the slope parameter $\tau$ from the fitting procedure for varying temperature $T$. The error bars indicate the statistical error (see text for details). }
  \label{fig:tau}
\end{figure}

If $C^{xy}(0)$ is one value of interest one gets out of the Green-Kubo formalism, the other one is the relaxation time $\tau$, the slope of the correlator, shown in \cref{fig:tau}.
One observes, that the $\tau$ parameter of the pion gas decreases smoothly and less rapid than that of the Hagedorn gas.
Here, no analytic estimator is available at the moment.

While during the fitting procedure, $C^{xy}(0)$ varies only little and agrees nearly perfectly with the analytic estimate, the results of the fits for the relaxation time $\tau$ vary drastically between different runs. Therefore also the statistical error of this quantity as obtained by calculating the Jackknife variance (for a review see \cite{Miller:1974}) is shown in \cref{fig:tau}.

Nevertheless, considering a relaxation time as the inverse of an interaction rate, one may express (in a low density approximation) $\tau=1/\Gamma=1/\langle n\sigma v_{\rm rel}\rangle$. Thus, the exponential behavior of $\tau$ as function of the temperature $T$ is mainly dictated by the increase of the particle density $n$.
It may be matter of debate, if the factor $\langle \sigma v_{\rm rel}\rangle$ really directly translates into the transport cross section $\sigma_{\rm tr}=2/3\sigma_{\rm tot}$. Here further investigations are at order.

It is very instructive to check the Green-Kubo method against some known input. For this, an implementation of the algorithm of generating random numbers with memory \cite{Schmidt:2014zpa} enables to dial in specific values for the correlation and compare with the results of the Green-Kubo method. Error estimates according a Jacknife method show clearly, that the error scales as usual with $1/\sqrt{N_{\rm run}}$, if $N_{\rm run}$ independent runs are performed, but with $1/N_{\rm timestep}$ in a single run. Thus it is more preferable, to perform long runs, than doing multiple short runs. In addition, having an estimate for the correlation time $\tau$, the effect of the timestep size may be estimated correctly.

\section{Results}

\begin{figure}[htb]
  \centering
  \includegraphics[width=\columnwidth]{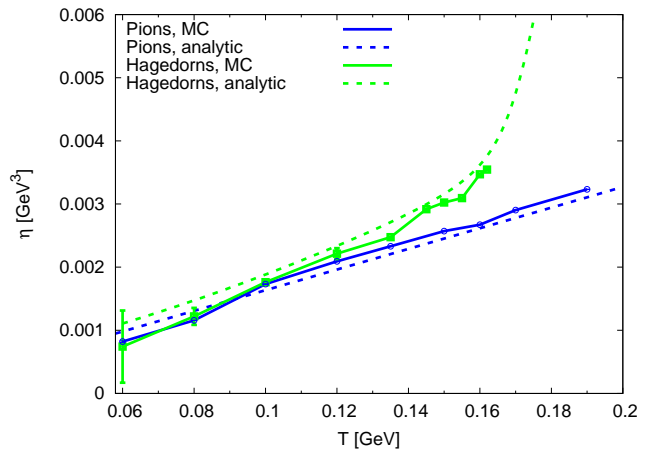}
  \caption{A comparison of analytical, \cref{eq:etachapens}, and numerical values of the shear viscosity $\eta$. The error bars shown emerge from the error bars of $\tau$ shown in \cref{fig:tau}. }
  \label{fig:eta}
\end{figure}

The final results for the shear viscosity using the Green-Kubo method \cref{eq:eta} are shown in \cref{fig:eta} and compared to analytic estimates.
The agreement is very well; while there is some tiny underestimation for $T<100\MeV$, Monte Carlo results coincide very well with the analytic estimates for higher temperatures.
Here one can also see, that $\eta$ stays more or less the same for both species at lower temperature and starts to diverge the more particles are created in the box in the Hagedorn case. The values explode, if the particle number density increases ad infinitum near $T_H$. Nevertheless, it increases less rapidly than the entropy density as shown in \cref{fig:s_T3}.

It is interesting to observe, that the competing differences in the intermediate result of $C(0)$ and $\tau$ cancel each other at low temperatures and only for $T\gtrsim 140\MeV$, a different behavior between the pion gas and the Hagedorn state gas my be observable.

\begin{figure}
  \centering
  \includegraphics[width=\columnwidth]{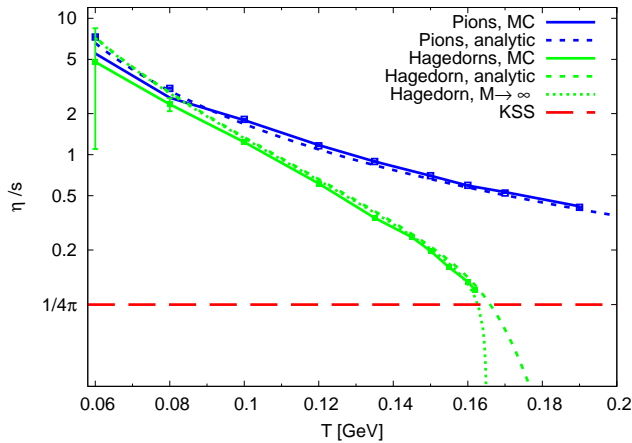}
  \caption{The final result is $\eta$ normalized to the entropy density for numerical and analytical estimates. Error bars are as in \cref{fig:eta}.
    The KSS bound $1/4\pi$ is indicated.
    The Hagedorn extrapolation $M\to\infty$ contains both separate $\eta$ and $s$ extrapolations.
  }
  \label{fig:eta_s}
\end{figure}

Combining both the results of the thermodynamical quantities (the entropy density), and the shear viscosity,
\cref{fig:eta_s} shows the final result, the shear viscosity to entropy density ratio.
As expected, at low temperatures the results for the pion gas and the Hagedorn gas coincide. Since the entropy density $s$ very rapidly starts to diverge with increasing temperature, also the fraction $\eta/s$ diverges. Finally, all the calculated results via the Monte Carlo/Green Kubo approach a stop at values above the KSS bound of $1/4\pi$. The analytic estimates indicate, that the results drop below this boundary and go to zero when temperature increases further.

Using the statistical error for $\tau$, one can compute the errors for $\eta$ and $\eta/s$. In \cref{fig:eta} and \cref{fig:eta_s} one observes, that the numerical results including the errorbars do not match the analytical curve.

This leads us to the finding, that there are some systematic error in the Green-Kubo formalism, which are underestimated in  the current work.

\section{Conclusions}
\label{sec:Conclusions}

In the present work, the transport coefficient $\eta/s$ has been calculated for a gas of Hagedorn resonances. Using the usual way of doing Monte Carlo simulations with a Green-Kubo analysis, it has been shown, that these results coincide very well with some analytic estimates. In addition, the same analysis has been performed for a single pion gas with elastic interactions. This, on one hand side allows to check the used analysis routines and also on the second hand side indicates the differences of the interactions.

Here, while the MC calculations only consider Hagedorn states with masses $m<10\GeV$, the analytic estimates allow to extrapolate to a Hagedorn spectrum up to infinite masses.
Interestingly, the influence of high mass Hagedorn states with $m>10\GeV$ is only visible in the present analysis at temperatures $T>160\MeV$, which are very close to the underlying Hagedorn temperature $T_H=165\MeV$.

Finally, the main result of this study is the finding, that the fraction $\eta/s$ drops while approaching the limiting Hagedorn temperature.
While $\eta$ itself increases with increasing temperature, the growth of $s$ overwhelms it and dominates the overall behavior.
The KSS bound is violated at $T_H$.

This singular behavior may be cured by a phase transition to some other phase with increasing $\eta/s$, being beyond the Hagedorn picture, since the Hagedorn temperature is a limiting temperature.

\begin{acknowledgments}
  The authors thank Harri Niemi for useful discussions.
  This work was supported by the Bundesministerium f\"ur Bildung und
  Forschung (BMBF), grant No.~3313040033.

\end{acknowledgments}


\bibliography{paper.bib}

\end{document}